\documentclass[twocolumn,showpacs,preprintnumbers,amsmath,amssymb,nofootinbib
,superscriptaddress]{revtex4}
\usepackage{hyperref}
\usepackage{graphicx}
\usepackage{color}
\usepackage{xcolor}

\newcommand{\abs}[1]{\left| #1 \right|}

\usepackage{diagbox} 
  
\def\beq{\begin{equation}}  
\def\eeq{\end{equation}}  
\def\bea{\begin{eqnarray}}  
\def\eea{\end{eqnarray}}

\def\bq{\begin{quote}}  
\def\eq{\end{quote}}

\def \lta {\mathrel{\vcenter  
     {\hbox{$<$}\nointerlineskip\hbox{$\sim$}}}}  
\def \gta {\mathrel{\vcenter  
     {\hbox{$>$}\nointerlineskip\hbox{$\sim$}}}}   

\relax

\begin{document}
\preprint{FERMILAB-PUB-19-603-T}

{\title{The Next Higgs Boson(s) and a Higgs-Yukawa Universality}

\author{Christopher T. Hill}\email{hill@fnal.gov}
\affiliation{Fermi National Accelerator Laboratory\\
P.O. Box 500, Batavia, Illinois 60510, USA}

\begin{abstract}
We consider multi-Higgs-doublet  models which, for symmetry reasons, have a universal Higgs-Yukawa (HY) coupling, $g$.
This is identified with the top quark  $g=g_t\approx 1$.
The models are concordant with the quasi-infrared fixed point, and the top quark mass
is correctly predicted with a compositeness scale (Landau pole) at $M_{planck}$, with
sensitivity to heavier Higgs states. The observed
Higgs boson is a $\bar{t}t$ composite, and a first sequential Higgs doublet, $H_b$, with $g\approx g_t\approx 1$ 
coupled to $\bar{b}_R(t,b)_L$
is predicted at a mass $3.0 \lesssim M_b \lesssim 5.5$ TeV and accessible to LHC and its upgrades. This would explain
the mass of the $b-$quark, and the tachyonic SM Higgs boson  mass$^2$.  The 
flavor texture problem is no longer associated with the HY
couplings, but rather is determined by the inverted multi-Higgs boson mass spectrum, e.g., the lightest
fermions are associated with heaviest Higgs bosons and vice versa. The theory is no
less technically natural than the standard model. The discovery of $H_b$ at the LHC would confirm 
the general compositeness idea of Higgs bosons and anticipate additional states potentially
accessible to the $100$ TeV $pp$ machine.
\\
\\
{Invited Talk\footnote{A video of this talk is available at:\\ 
www.perimeterinstitute.ca/conferences/simplicity-iii \\
in ``videos,'' title:  ``Where are the Next Higgs Bosons?''}  delivered at
{\bf \em Simplicity III}, Perimeter Institute, Waterloo, Ontario, Canada, Sept. 11, 2019}
\end{abstract}

\maketitle
 
\date{\today}

\section{Introduction: Predicting $m_{top}$}

In 1980 the earliest studies of the renormalization group (RG) flow
of the top quark Higgs-Yukawa coupling 
were undertaken \cite{PR}\cite{FP}.  The skeletal RG equation
for  the top quark Higgs-Yukawa coupling, $g_t$, is:
\beq
\label{one}
D g_t = g_t\left(\left(N_c+\frac{3}{2}\right)g_t^2 - \left(N_c^2-1\right)g_3^2\right)
\eeq
where $D= 16\pi^2 \partial/\partial \ln(\mu)$,  $\mu$ is
the running mass scale,  $g_3$ is the QCD coupling, and $N_c=3$ is the number of colors.
For illustrative purposes we discuss the one-loop RG equation and suppress electroweak corrections.

Starting the running of $g_t(\mu)$ at very
large mass scales, $\mu = M_X$, with large initial values, i.e., $g_t(M_X)>>1$
(effectively a Landau pole at $M_X$), 
it is seen that $g_t(\mu)$  flows into an ``infrared 
 quasi-fixed point.''
This is ``quasi'' in the sense that, 
if the QCD coupling, $g_3$, was a constant
then $g_t$ would flow to an exact conformal fixed point.  
The  low energy prediction of the top quark  Higgs-Yukawa (HY) coupling from the infrared quasi-fixed point
is very insensitive to its precise, large initial
values and mass scales. This is often nowadays called a ``focus point,''
since a focusing occurs
in the infrared, and the top quark mass
is determined to lie within a narrow range of values.  

This is shown in Fig.(1)
where the effective top mass $m_{top}=g_t(\mu) v $  (where $v = 175$ GeV is the electroweak scale)
is plotted vs. renormalization scale $\mu$ (GeV); the physical top mass corresponds $\log(\mu)\sim 2$.
Assuming the initial mass scale of the Landau pole (LP) is of order $10^{15}-10^{19}$ GeV, 
the quasi-fixed point translates 
into values of the top quark mass that are heavy, as seen in Table I. 
One obtains $m_{top} \sim 220$ GeV starting with the LP at the Planck mass.

\begin{figure}[t!]
\vspace{-1 in}
	\includegraphics[width=0.5\textwidth]{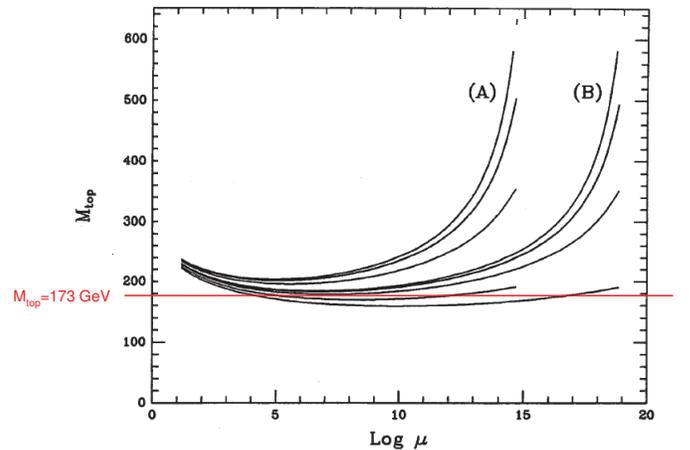}
	\vspace{-1.2 in}
	\caption{The top quark quasi--fixed point in the top mass  $m_{top}=g_tv$ where $v=175$ GeV
	is the Higgs VEV, plotted vs the running scale $\mu$, The focusing in the infrared
	and its relative insensitivity to initial values is indicated.}
	\label{RGFlowgtopSM}
\end{figure}
Initially this prediction for an ultra-heavy top mass was greeted with some derision. The
favored top quark mass values in the early 1980's ranged from $\sim 15$ GeV to $\sim 26$ GeV 
and it was evidently considered absurd
that an elementary fermion in the standard model (SM) would have
a mass comparable to the weak scale.
However, as the 
experimental and indirect lower limits for $m_{top}$
climbed above $\sim 100$ GeV, the possibility 
of the ultra-heavy top quark became a high probability.

\begin{table}
\begin{center}
\renewcommand{\arraystretch}{2.0}
    \caption{Top quark mass for initial Landau pole $M_X$.}
    \label{table1}
    \vspace{0.05 in}
    \begin{tabular}{l|c|c|c|c|c} 
   $M_X$ GeV & $10^{19}$  & $10^{15}$  & $10^{11}$  & $10^{7}$  & $10^5$ \\
      \hline
      $m_{top}$ GeV & 220  & 230 & 250 & 290 & 360 \\
    \end{tabular}
  \end{center}
  \vspace{-0.1 in}
\end{table}

Prior to the top quark discovery,
following suggestions of Nambu, \cite{Nambu}, and pioneering work of Miransky, Tanabashi and
Yamawaki, \cite{Yama},  a predictive theory of a composite Higgs boson  
was constructed by Bardeen, Hill and Lindner \cite{BHL}.
This is based upon a Nambu-Jona-Lasinio model \cite{NJL} where
the Higgs is composed of $\bar{t}t$ \cite{review}. 
It turns out that the solution to the top condensation theory is, in fact, the RG quasi-fixed point! 
The quartic coupling of the Higgs, $\lambda(\mu)$, is also governed by a similar infrared quasi-fixed point 
(see the detailed RG discussion in \cite{BardeenHill}).

Ultimately, the top quark was discovered in 1995 at Fermilab by 
CDF and D-Zero \cite{CDF}\cite{Dzero}, with a mass of $m_t=174$ GeV.
 The top mass 
is shy of the naive SM infrared quasi-fixed point 
with $M_X=10^{19}$ GeV by about 20\%.
I consider this to be a success of the infrared quasi-fixed point, 
given that the prediction assumed only SM
physics extending all the way up to Planck scale.  
To me this may be indicative of a composite Higgs boson of $\bar{t}t$,
but the whole story is yet to be unraveled.

I have long wondered what would convincingly bring the top mass quasi-fixed point prediction
into concordance with experiment.  It turns out that it isn't easy to do this with
minor modifications of the SM.  For example,
one might consider imbedding $SU(3)\rightarrow SU(3)\times SU(3)...$,
at intermediate scales in the RG running,
such as a perturbative version of ``topcolor,'' or ``flavor universal colorons,'' or 
``latticized extra dimensions.''  However, this
generally causes the effective $g_3$ to become larger and the quasi-fixed point 
of $g_t(m_{t})$ moves up and 
the discrepancy with experiment gets larger.
Moreover, in the $\bar{t}t$ composite  scheme the quartic Higgs coupling is too
high $\lambda \sim 1$ vs. the SM value, $\lambda\sim 0.25$,
and leads to an unacceptably large Higgs boson mass \cite{BHL}. 
This may also be informing us of required modifications of the renormalization group
beyond the simple SM inputs at higher energy scales to
bring the result into concordance with experiment \cite{HillThomsen}.

\vspace{0.1 in}

\section{Scalar Democracy}

Recently we have considered a rather drastic, 
maximal scalar field extension of the SM \cite{HMTT} (and a less drastic
minimal version, \cite{HMTT2}).
This was largely motivated by curiosity: How obstructive are the rare weak decay flavor constraints 
on a rich spectrum of Higgs bosons? How many scalars might exist, given the fermion
composition of the SM, and what patterns might be suggested, etc.?

We call this model ``Scalar Democracy.''  We count the allowed vertex operators
(see Section III) and we find that
this leads to a vast spectrum of color octet isodoublets
and triplet leptoquarks, etc.  
We assume these exotic new scalar bosons have ultra-large  masses
and do not affect low energy RG running.
 The model also contains eighteen Higgs doublets in the  quark sector
and likewise in the lepton sector that will have  masses
extending up to $\sim 10^5$ TeV.  These two subsectors resemble an $SU(6)_L\times SU(6)_R$
linear $\Sigma$-model Lagrangian \cite{manohar} (as in Fig.(2)), where the interaction is subcritical and ultimately
only the SM Higgs condenses. 

The theory has one universal HY coupling $g$ defined
at the Planck scale, which we view as the compositeness scale of all scalars.
This universal $g$ is renormalized as we flow into the infra-red  and symmetry breaking
effects lead to slightly different values in the different subsectors of the theory.
For quarks this coupling near the weak scale is identified with the top quark, $g_t=g$, 
while for leptons we obtain $g_\ell \sim 0.7 g$. 
Hence, our theory is calibrated by the known value of $g_t\approx 1$ \cite{HMTT}.
This buys us some predictivity. 

\begin{figure}[t]
\vspace{-2.2 in}
	\hspace*{-4.5cm}
	\includegraphics[angle=-90,width=1.0\textwidth]{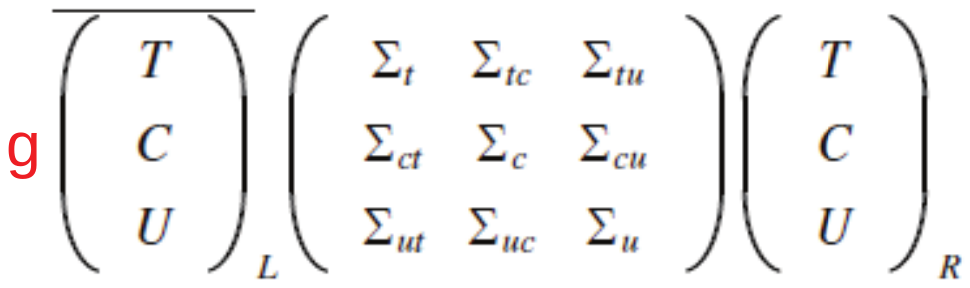}
	 \vspace{-2.0 in}
	 \vspace{-0.5 cm}
	 \caption{ 
	 $SU(6)_L\times SU(6)_R$ chiral Lagrangian structure of the scalar democracy 
	 in the quark sector.  $T=(t,b)$, $C=(c,s)$, $U=(u,d)$ and $\Sigma_{xy}=(H_x, H_y^C)$.
	The quark Higgs subsector becomes a subcritical $SU(6)_L\times SU(6)_R$
linear $\Sigma$-model Lagrangian, where $\Sigma$ is a $6\times 6 $ complex matrix that can be viewed
as nine $2\times 2$ complex $\Sigma$ fields where each $\Sigma_{xy}=(H_x, H^C_y)$
is a pair of Higgs doublets. There is one universal coupling $g$ which is subject to
RG effects.}
	\label{fig:Ft_regions}
\end{figure}

Here
 the  flavor physics and mass hierarchy problems are flipped out of the Higgs-Yukawa coupling texture (the
texture is now that of Fig.(2) with the single universal coupling $g$) and into
the mass matrix of the many Higgs fields. 
The Higgs mass matrix is input as $d=2$ gauge invariant operators. We have no theory
of these, but we must choose the inputs to fit the quark and lepton sector masses
and CKM physics, as  well as maintain consistency with rare weak decays, etc.  
It is not obvious {\em a priori} that there exists a consistent solution with the flavor 
constraints. 
 
If we zoom in on the  top-bottom, upper left corner of Fig.(2),
we have $\Sigma_{t}=(H_t, H^C_b)$  where $H_t=H_0$ is the SM Higgs doublet
with coupling $g$ to $T_Lt_R$ and $H_b$ is a new isodoublet, also with coupling $g$ 
to $T_Lb_r$. $H_0$ will condense, while mixing of $H_b$ with $H_0$ gives $H_b$ a small ``tadpole'' VEV
and accounts for the $b$ quark mass. This also back-reacts, generating the tachyonic $H_0$ mass and VEV.  
This mixing is generalized  throughout the entire sector and generates the masses and mixing angles of quarks.
The third generation subsector model can also be viewed as a self-contained theory \cite{HMTT2}
and is discussed in Section IV.


The RG equation for the universal
HY coupling  in the quark subsector now takes the one-loop form
\cite{HMTT}\cite{HillThomsen}:
\beq
\label{two}
D g_t = g_t\left((N_c+N_f)g_t^2 - (N_c^2 -1)g_3^2\right)
\eeq
Note the enhanced coefficient of $g_t^2$ in eq.(\ref{two})
relative to eq.(1) where $N_f$ is the number
of flavors, i.e., $N_f=6$ in the quark sector of the model.

\begin{figure}[t!]
\vspace{-1 in}
	\includegraphics[width=0.5\textwidth]{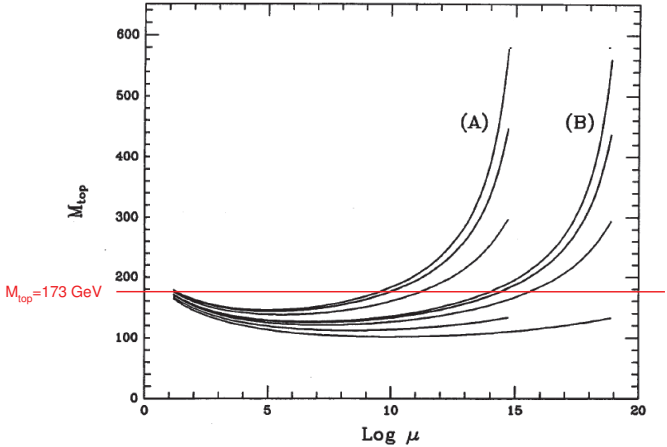}
	\vspace{-1.2 in}
	\caption{The top quark fixed point is shifted down in the $SU(6)\times SU(6)$
	model and becomes concordant with the observed top mass, with residual
	corrections coming from sensitivity to the extended Higgs spectrum (seee Table II). }
\end{figure}

\begin{table}
\begin{center}
\renewcommand{\arraystretch}{2.0}
    \caption{Top quark mass for initial Landau pole $M_X$.}
    \label{table2}
    \vspace{0.05 in}
    \begin{tabular}{l|c|c|c|c|c} 
   $M_X$ GeV & $10^{19}$  & $10^{15}$  & $10^{11}$  & $10^{7}$  & $10^5$ \\
      \hline
      $m_{top}$ GeV & $156+\Delta$  & $162+\Delta$ & $177+\Delta$ & $205+\Delta$ & $254+\Delta$ \\ 
    \end{tabular}
  \end{center}
  \vspace{-0.1 in}
\end{table}

The resulting RG evolution is shown in Fig.(3) and the top quark mass predictions are given in Table II.
The residual sensitivity to the decoupling of the heavy Higgs bosons 
reduces the effective $N_f$ below the Higgs decoupling scale.  We estimate that $\Delta=
(2.8\;\makebox{GeV})\ln(\langle M_H \rangle/10^2)$ hence $\Delta \approx 19$ GeV for an average
heavy doublet mass $ M_H \approx 100$ TeV.  Hence, the top mass is correctly
predicted for $M_X \sim 10^{19}$ GeV, and is indirectly probing the new physics of the additional
heavy Higgs bosons.

\begin{table*}
\label{diagonal}
\centering
\renewcommand{\arraystretch}{1.3}
\begin{tabular}{|p{3cm}p{4.1cm}p{3cm}l|} 
	\hline
	Higgs field & Fermion mass &  Case (1) [TeV]  &  Case (2) [TeV] \\
	\hline
	\hline
        $H^{\prime}_0 = v + \frac{h}{\sqrt{2}}$ & $m_t = gv=175$ GeV  & $m_H=0.125$  &  $m_H=0.125$ \\
	\hline
	$H^{\prime}_{b}=v\frac{\mu^2}{M_b^2} + H_{b}$ & $ m_b= gv\frac{\mu^2}{M_b^2}=4.5$ GeV   & $M_b=3.5$  &  $M_b=0.620  $     \\
	\hline
	$H^{\prime}_{\tau}=v\frac{\mu^2}{M_\tau^2} + H_{\tau}$ &  $ m_\tau= g_\ell v\frac{\mu^2}{M_\tau^2}=1.8$
	GeV &  $M_\tau= 6.8 $   &  $M_\tau= 0.825 $  \\
	\hline
	$H^{\prime}_{c}=v\frac{\mu^2}{M_b^2} + H_{c}$ &  $ m_c= gv\frac{\mu^2}{M_c^2}=1.3$ GeV &  $M_c=13.5$  &  $M_c=1.2$  \\
	\hline
	$H^{\prime}_{\mu}= v\frac{\mu^2}{M_\mu^2} + H_{\mu}$ &  $ m_\mu= g_\ell v\frac{\mu^2}{M_\mu^2}
	=106$ MeV &  $M_\mu= 1.2\times 10^2$  & $M_\mu= 3.4$   \\
	\hline
	$H^{\prime}_{s}=v\frac{\mu^2}{M_s^2} + H_{s} $ & $m_s= gv\frac{\mu^2}{M_s^2}=95$ MeV & $M_s= 1.8\times 10^2$  & $M_s= 4.3 $  \\
	\hline
	$H^{\prime}_{d}=v\frac{\mu^2}{M_d^2} + H_{d} $ &  $m_d= gv\frac{\mu^2}{M_d^2}=4.8$ MeV 
	& $M_d=3.6\times 10^3  $ & $M_d=19$  \\
	\hline
	$H^{\prime}_{u}=v\frac{\mu^2}{M_u^2} + H_{u} $ &  $m_u= gv\frac{\mu^2}{M_u^2}=2.3$ MeV 
	& $M_u= 7.6\times 10^3 $  & $M_u= 27 $ \\
	\hline
	$H^{\prime}_{e}= v\frac{\mu^2}{M_e^2} + H_{e}$ &   $m_e= g_\ell v\frac{\mu^2}{M_e^2}=0.5$ MeV
	& $M_e=2.45\times 10^4$  &  $M_e=49$ \\
	\hline
\end{tabular}
\caption{The estimates for diagonal Higgs bosons masses 
in the limit of no CKM mixing, and assuming 
(1) the level-repulsion feedback on the 
Higgs mass term is limited to $(100 \,\text{GeV})^2$  for each of the quarks and leptons, hence $M_q= (100$ GeV$)(m_t/m_q )$ and 
$M_\ell= (100$ GeV$)(g_{\ell}m_t/m_{\ell})$.
(2) $\mu=100$ GeV for all mixings, hence $M_q= \mu (m_{t}/ m_{q})^{1/2}$  and $M_\ell= \mu (m_{t}g_{\ell}/gm_{q})^{1/2}$.
Here  $g=1$, $g_\ell=0.7$ and $ v= 175 $ GeV. For a more detailed discussion see \cite{HMTT}. 
}
\end{table*}

\vspace{0.2 in}
 
 The CKM physics is generated by the off-diagonal Higgs fields and their mixings.
We ran numerous benchmark tests of hypothetical Higgs mass matrices to
study consistency with flavor physics (rare weak decay constraints)
and the generation of quark and lepton masses. A sample of these are displayed in  Fig.(4).
The results are nontrivial, yet
{\em we find there are consistent solutions where the
masses and mixings of a spectrum of new Higgs doublets can explain the
entire fermionic mass and CKM-mixing structure of the SM
with the single universal HY coupling.}

\begin{figure*}
   	\centering
   	\includegraphics[width=.95\textwidth]{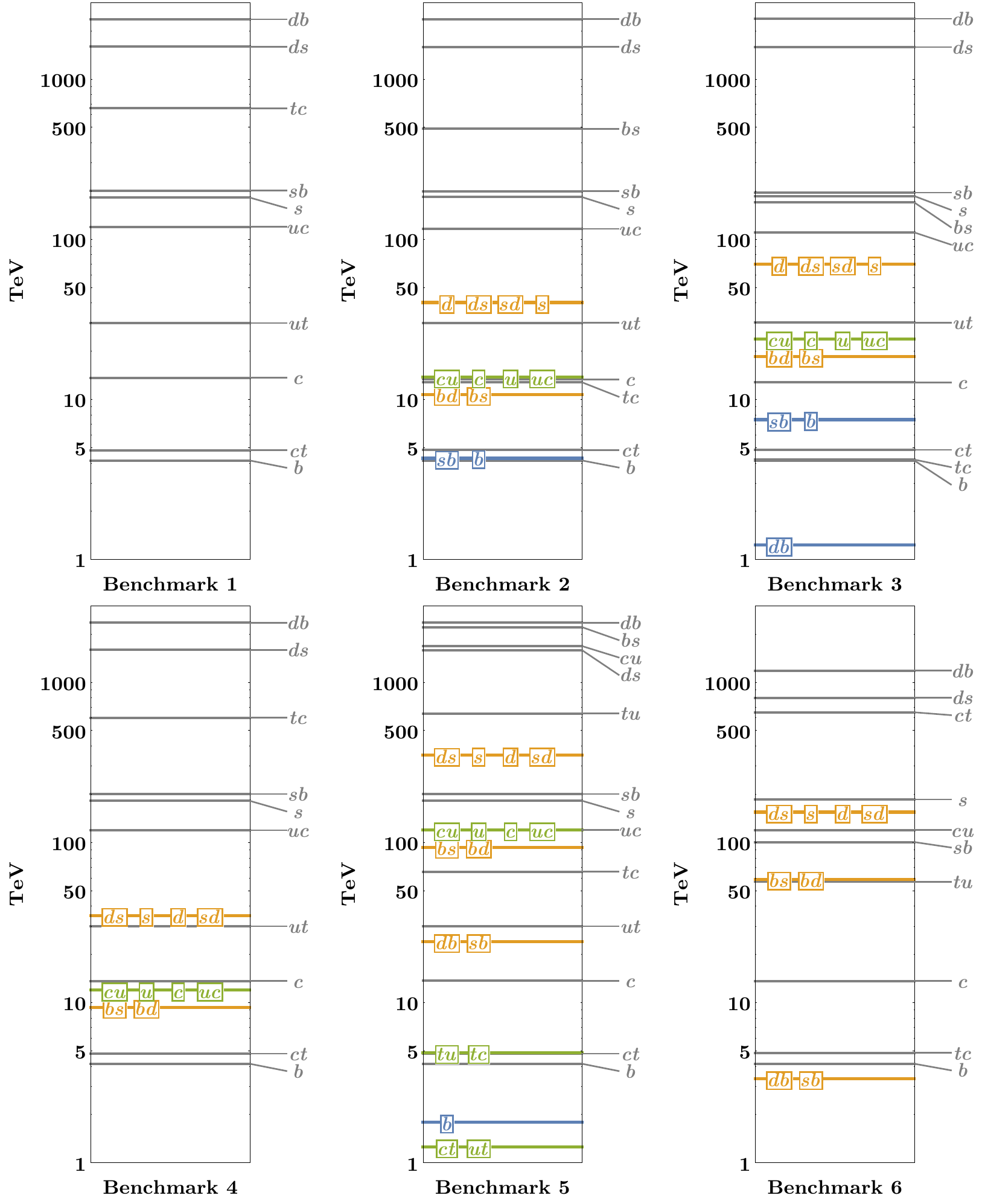}
   	\caption{Experimental constraints and mass estimates for the heavy Higgses $H_{xy}$ in the quark sector in six different benchmarks. The labels denote the indices of the corresponding Higgs. The gray lines are the Higgs mass estimates. The colored lines correspond to the most stringent experimental lower bounds on each of the Higgs masses: green if the constraint is from $ D_0 $ mixing, orange from $ K_0 $ and blue from $ B_s $. If a mass-estimate entry is not shown, it is above the scale of the plot. Similarly if a mass bound is not shown it is below the scale of the plot.  Hence we see that Benchmarks 3 and 5 represent failures, since the  $H_c$ mass (grayline c) lies below the green lower bound for $H_c$, whereas Benchmark 4 is a success. See \cite{HMTT} for details.}
   	\label{fig:meson_constraints}
   \end{figure*}

Here is a brief summary of the benchmarks, but we 
must refer the reader to \cite{HMTT} for more detail:  
 
\emph{Benchmark 1} shows the mass spectrum of the theory in the absence of right-handed rotation. 
The Higgs mass estimates exhibit the inverse hierarchy
with the Higgses associated to the lighter quarks tending to be the heaviest. 
A non-trivial left-handed rotation will lower the mass estimates for the off-diagonal mass. 
In particular, $ H_{ct} $, must be almost as light as $H_b $ in order 
to generate a large off diagonal mass element in the fermion mass matrix. 
The plot does not show bounds on the Higgs masses, 
as we have not included the loop-induced effects here. 

\emph{Benchmarks 2 and 3} illustrate a marginal case, and how much right-handed rotation 
can be allowed without tension between mass estimates and flavor bounds.  

\emph{Benchmarks 4 and 5}  parametrize mixing in terms of powers of the CKM matrix, 
to illustrate how large the right-handed mixings can be in terms of a more familiar matrix.  

\emph{Benchmark 6} puts the entire mixing into a left-handed rotation 
and gives an example of evading the flavor bounds.  
The main constraints in this case are from $K^0$ mixing.

We also show, in Table III, the diagonal Higgs masses in the limit of no CKM mixing with various
assumptions on the heavy Higgs mass mixing with the SM Higgs \cite{HMTT}.
  
In this theory we didn't consider in detail the constraint of grand unification, e.g., $\sin^2\theta_W$
predictions.  I view unification as something we might ultimately retrieve, however, to
have the simplest unification, such as $SU(5)$  already requires 
further extension of the scalar sector 
since the $\bf 24$ is not contained in our extended spectrum, while multiplets
of the scalar $\bf 5$ are present.  Eight Higgs doublets can easily unify in the SM \cite{Harada}. Beyond
this it appears  that to have couping constant unification may require the {\em ad hoc} addition
of more color states that don't interfere with the running of $g_t$. 
The final story here may be significantly different than the conventional one,
involving perhaps
large Planck scale effects, exotic unification, and perhaps composite gauge fields.

The generic idea of a large spectrum of Higgs bosons is not without precedent, and  we only briefly
mention some of the literature.  The ``Private Higgs''
models of BenTov, Porto, and Zee, et.al., \cite{Zee}\cite{Bentov},
have diagonal Higgs fields (typically $6$ in each subsector) and 
expect new Higgs states in the TeV mass range, while CKM mixing is input into the model by hand.
Related enhanced spectroscopy of composite Higgs bosons arises in various
other arenas, e.g.,
\cite{Marco}\cite{Osipov}\cite{Rode}\cite{Ishida}.
Bjorken inspired our active thinking about this in private discussions
of multi-scalars in extra-dimensional gravity models.
A similar large number of Higgs doublets
has been invoked in an interesting  model of Weinberg's asymptotic safety \cite{San}.
This may preclude a conventional GUT picture if the unification of gauge interactions
occurs in tandem with the non-perturbative asymptotically safe fixed point.

Scalars go hand-in-hand with Weyl invariance. Recent  models
suggest a unified picture of inflation occuring with the dynamical generation of
the Planck scale \cite{Weyl}, which may provide a larger context for this scenario.
Also, the idea of universal large Higgs-Yukawa coupling, e.g.,  $H_b$ with $g=1$, 
was previously introduced in the context of a Coleman-Weinberg Higgs potential \cite{CWH}
(see also \cite{Rao}\cite{thesis}).

\section{Counting Scalars }
 
The fixed point of the HY coupling is intimately intertwined with
compositeness of Higgs bosons. In the case of top condensation, if we renormalize the 
Higgs field so the HY coupling becomes $\bar{\psi}_L t_R H$, fixed to unity, the
coupling then appears in the Higgs kinetic term, $(1/g_t^2) DH^\dagger DH$.  
A Landau pole, $g_t\rightarrow \infty$,  then implies the Higgs kinetic term vanishes
at the composite scale, and at that scale the Higgs becomes an ``auxiliary field.''  
We can then integrate the Higgs bosons out of the theory and 
replace the HY interactions with  four fermion interactions, i.e., a Nambu-Jona-Lasinio
theory. Above the scale of the LP, we expect
that this four fermion interaction is generated by a new gauge force, or gravity.
It is the matching of the four fermion
theory at the Landau pole to  auxiliary Higgs fields that defines the
composite Higgs bosons.  

All of the SM matter fields can be represented by $48$ two-component left-handed 
spinors, $\psi _{A}^{i}$. 
This includes all the left-handed and anti-right-handed fermions.
We can collect these into a large global $SU(48)\times U(1) $ multiplet, 
the new dynamics that is blind to the SM gauge interactions.
We emphasize that this is a dynamical symmetry, and familiar GUT 
theories that contain only the SM fermions will be gauged subgroups of this $SU(48)$.
Here the indices $\left( i,j\right) $ run over all the $48$ flavor, doublet, 
and color degrees of freedom of the SM fermions.

The most general non-derivative ($s$-wave) scalar-field  bilinear 
we can construct of these fields is the vertex operator, $\epsilon^{AB}\psi_{A}^{i}\psi_{B}^{j}$,
which induces a composite field $\Theta_{ij} $ as:
\bea
\epsilon^{AB}\psi_{A}^{i}\psi_{B}^{j}\Theta_{ij}+h.c.,
\eea
where $\Theta _{ij}$ transforms as the symmetric $ \mathbf{1176} $ representation of  $SU(48)$ 
(this is analogue to the sextet representation of $ SU(3) $). 
The field $\Theta _{ij}$ contains many complex scalar fields with assorted quantum numbers, 
including baryon and lepton number, color, and weak charges.

To make contact with the SM fields, we consider the usual $24$ 
left-handed quarks and leptons, $\Psi_{Li}$, and the $24$ right-handed counterparts, $\Psi_{R\widehat{i}} $.
 The index $i$ now runs over the chiral $SU(24)_{L}$ and $\widehat{i}$ over the chiral $SU(24)_{R} $ subgroups of $ SU(48) $. We thus have:
\bea
\label{fields}
\Phi_{i\widehat{j}}\overline{\Psi }_{L}^{i}\Psi _{R}^{\widehat{j}}+\Omega
_{ij}\overline{\Psi }_{L}^{i}\Psi _{R}^{jC}+\widehat{\Omega }_{\widehat{ij}%
}\overline{\Psi }_{R}^{\widehat{i}}\Psi _{L}^{\widehat{j}C}+\text{h.c}.,
\eea
where $ \Phi_{i\widehat{j}} $ is the $ (\mathbf{24}_L,\, \mathbf{24}_R) $ 
complex scalar field with $24^{2}=576$ complex degrees of freedom. 
$\Omega $ and $ \widehat{\Omega } $ are the symmetric $\mathbf{300} $ representation of 
$SU(24)_{L}$ and $SU(24)_{R}$ respectively.
 Together these match  the degrees of freedom of $\Theta _{ij}$.
Here $\Omega _{ij}$ and $\widehat{\Omega }_{ij}$ are the analogues of Majorana masses and carry fermion number, while $\Phi $ contains fermion number neutral fields, such as Higgs fields, in addition to ($B-L$) leptoquark multiplets.

The $\Phi ,\Omega$ and $\widehat{\Omega }$ fields can be viewed as the ``composite fields'' 
arising from a NJL model effective description of the new forces.
Consider just the $SU(24)_{L} \times SU(24)_{R}\times U(1) \times U(1)_{A}$ invariant NJL model:
\bea
\label{eq:NJL}
- \frac{g^{2}}{M^{2}} (\overline{\Psi}_{L}^{i} \Psi _{R}^{j}) (\overline{\Psi}_{R,i} \Psi_{L,j}),
\eea
where the negative sign denotes an attractive interaction in the potential.
It should be noted that we can equally well write current-current (and tensor-tensor) 
interactions, mediated by 
heavy spin-1 bosons (or Pauli-Fierz spin-2 gravitons); 
these will generally contain scalar
channels and will Fierz rearrange to effectively reduce to eq.(\ref{eq:NJL}) with the attractive signs.
There also exists the possibility of the following NJL models:
\bea
-\frac{g^{2}}{M^{2}} (\overline{%
\Psi }_{L}^{i} \Psi_{R}^{Cj} ) (\overline{\Psi}_{R,i}^{C} \Psi_{L,j} ) \;\;
\makebox{or} \;\; \left( R \leftrightarrow L\right),
\eea
which lead to the composite bosons $\Omega $ and $\label{tab:Higgs_masses}\widehat{\Omega}$. 
The first step to solving an NJL theory would be to factorize the interaction of eq.(\ref{eq:NJL}) by introducing auxiliary scalar fields.
This leads to the equation we started with,  eq.(\ref{fields}), where  $\Phi ,\Omega$ and $\widehat{\Omega} $ are auxiliary fields.

A universal flavor and color blind interaction will bind fermion pairs into scalars 
that are bound states of ordinary quarks and leptons and will generate a plethora of Higgs doublets. 
These bound states will have a universal Yukawa coupling $g$ at the scale $M^{2}$.
Moreover, with $g$ taking on a near-but-subcritical value, these bound states will 
generally have large positive masses but can be tuned to be lighter than $M$.
Symmetry-breaking effects are required to split the spectroscopy, including the 
SMH down to its observed negative mass term.
All other doublets remain heavy, but will mix. 

If $ g $ is supercritical then some or all multiplets will acquire negative renormalized masses, 
$ M^{2}<0 $ and the theory develops an overall vacuum instability.
For example, the  field  $\Phi ^{ij}$ with a supercritical coupling 
will generally condense into a diagonal VEV, $\left\langle \Phi
_{ij} \right\rangle = V \delta _{ij}$. This would become a spontaneously broken 
$\Sigma $-model of $ SU(24)_{L} \times SU(24)_{R} \times U(1) \times U(1)_A $.
In this supercritical case, all the fermions would acquire large, 
diagonal constituent masses of order $gV$, grossly
inconsistent with observation.

However, the structure we have just outlined can be subcritical. It will then 
contain many composite Higgs doublets with a spectrum of positive $M^2$'s  
and all fermions would be massless. 
Exactly how the scalar mass spectrum is generated is beyond the scope of 
our present discussion.
It can come from a scale invariant theory with extended quartc interactons and additional scalars.
This in fact could connect up to Weyl invariant theories in which even the Planck scale 
is generated spontaneously \cite{Weyl}, and will be intimately intertwined with inflation.

We assume such a spectrum of masses and mixings between the bound 
state scalars  that allows for a light sector from the SMH  to multi-TeV scales 
exists and extends up to the highest scales.  
We  assume that  $\Omega _{ij}$, $\widehat{\Omega }_{ij}$ and all color-carrying weak doublets
have very large positive $M^{2}$ and therefore we will ignore them. They will be inactive
in the RG evolution (though they may be welcome when unification is included).
There is fine-tuning required to engineer the light $M^2$ as $d=2$ operators, however many
of these terms are technically natural, protected by the $SU(48)$ symmetry structure. In
a subset model presented below, keeping only four Higgs fields for the third generation, we see explicitly that
the theory is no less techncally natural than the SM with only one unnatural fine-tuning.

Let us examine the quantum numbers of the spectrum of states in the $\Phi ^{ij}$ system.
Here we have:
\begin{itemize}
\item $ 9 \times (\mathbf{1}, \, \mathbf{2},\, \tfrac{1}{2}) \sim \bar{Q}_{L} U_{R} $; $ 3^2 \times 1\times 2 = 18$ complex degrees of freedom (DoFs),

\item $ 9 \times (\mathbf{1}, \, \mathbf{2},\, -\tfrac{1}{2}) \sim \bar{Q}_{L} D_{R} $; $ 3^2 \times 1\times 2 = 18$ complex DoFs,

\item $ 9 \times (\mathbf{1}, \, \mathbf{2},\, \tfrac{1}{2}) \sim \bar{L}_{L} N_{R} $ leptonic; $ 3^2 \times 1\times 2 = 18$ complex DoFs,

\item $ 9 \times (\mathbf{1}, \, \mathbf{2},\, -\tfrac{1}{2}) \sim \bar{L}_{L} E_{R} $ leptonic; $ 3^2 \times 1\times 2 = 18$ complex DoFs,

\item $ 9 \times (\mathbf{8},\, \mathbf{2},\, \pm \tfrac{1}{2}) \sim \bar{Q}_L \lambda^{a} U_R [D_R] $; $ 3^2 \times 8 \times 2 \times 2 = 288$ complex DoFs, 

\item $ 9 \times (\mathbf{3},\, \mathbf{2},\, \tfrac{1}{6} [-\tfrac{5}{6}] ) \sim \bar{L}_L  U_R [D_R] $; $ 3^2 \times 3 \times 2 \times 2 = 108$ complex DoFs,

\item $ 9 \times (\bar{\mathbf{3}},\, \mathbf{2},\, -\tfrac{1}{6} [-\tfrac{7}{6}] ) \sim \bar{Q}_L N_R [E_R] $; $ 3^2 \times 3 \times 2 \times 2 = 108$ complex DoFs,
\end{itemize}
where the brackets denote the SM quantum numbers. 
The first four entries in the above list are the $36$ Higgs doublets in the quark and lepton
sectors respectively.

To get an idea of the spectrum of heavy Higgses in this theory we consider the no-CKM-mixing
limit of the theory and quote the masses of the diagonal Higgs fields 
in Table III (see\cite{HMTT}).

\section{The Top-Bottom Subsector}

What might convince us that this system exists in nature?  We think 
the discovery of the first sequential Higgs boson, the $H_b$ with
a coupling $g\approx g_t\sim O(1)$ to $\bar{b}b$ would 
lend compelling credibility to the scenario.
As we will see, the $H_b$  has an upper bound
on its mass of about $5.5$ TeV. It may be
discovered up to $\sim 3.5$ TeV with the LHC, and conclusively so with the energy doubled LHC.  
Indeed, the LHC aleady has the capability of placing useful
limits on the $H_b$.

The top-bottom subsystem is a subsector of the full scalar democracy
which can be defined in a self-contained way.  It can also be
easily extended to include the $(\tau, \nu_\tau)$ leptons, and we will only quore results. 
The model as such is discussed in \cite{HMTT2}.
We presently assume the top-bottom subsystem is approximately
invariant under a simple extension of the Standard Model symmetry group
structure
\beq
\label{G}
G = SU(2)_L\times SU(2)_R\times U(1)_{B-L}\times U(1)_A.
\eeq
By ``approximately''  we mean that, if we  turn off
the $U(1)_Y$ gauging, $g_1\rightarrow 0$,
the symmetry $G$ is exact in the $d=4$ operators
(kinetic terms, Higgs-Yukawa couplings, and potential terms).
The SM gauging is the usual $ SU(2)_L \times U(1)_Y $, and 
is a subgroup of $G$.
The $U(1)_Y$ generator is now $I_{3R} + (B-L)/2$.
This electroweak gauging weakly breaks 
the symmetry $ SU(2)_R \times U(1)_{B-L} \rightarrow U(1)_Y \times U(1)_{B-L}$,
however, the $SU(2)_R$ remains as an approximate global symmetry of the $d=4$ operators.
In addition,
a global $U(1)_A$ arises as well.  $G$ provides custodial symmetry
so that  new symmetry breaking effects, in $d=2$ operators, are technically natural.  
Beyond the usual naturalness issue
of the Higgs boson mass, our present model is no less natural than the Standard Model.

To implement $G$ in
the $(t,b)$ sector we require that the SM Higgs doublet,
$H_0$, couples to $t_R$ with coupling  $y_t\equiv g_t$ in the usual way, and a second Higgs doublet, $H_b$,
couples to $b_R$ with coupling $g_b$. The symmetry $G$ then dictates 
that there is only a single 
Higgs-Yukawa coupling $g= g_t =g_b$ 
in the quark sector.  This coupling is thus determined by
the known top quark Higgs-Yukawa coupling, $y_t =g \simeq 1$.

Since $m_b/m_t \ll 1$,  the $SU(2)_R$ must be broken. 
Here we deploy ``soft'' symmetry breaking
through bosonic mass terms which
preserves the universality of quark Higgs-Yukawa couplings $g$.
Here we have three {\em a priori} unknown renormalized mass parameters,
$M_H^2$ (the input SM Higgs potential mass), $\mu^2$ (the mixing of $H_b$ with the 
SM Higgs)
and $M_b^2$ (the heavy $H_b$ mass).
Once
these are specified, we obtain the $b$-quark mass 
$m_b\sim m_t (\mu^2/M_b^2)$, and the
Higgs potential mass, 
$M_0^2=-({88.4}\;\makebox{GeV} )^2$ $=M_H^2 -\mu^4/M_b^2$  (note: $\abs{M_0} =
{125}\;\makebox{GeV} /\sqrt{2}$).
The tachyonic (negative) $M_0^2$ can 
be generated from an unknown input value, $M_H^2$. Here {\em we should not
fine tune the difference},  $M_H^2 -\mu^4/M^2$.  This requirement
establishes the scale of $M_b \sim {5}$ TeV.
Or, if we postulate that the input Higgs mass is small, $M_H^2 \ll M_b^2$, we immediately obtain $M_b \simeq {5.5}$ TeV.

The assumption of the symmetry, $G$, of Eq.~(\ref{G}) for the top-bottom system
leads to a Higgs-Yukawa (HY)
structure that is reminiscent of  the chiral Lagrangian of the proton and neutron
(or the chiral constituent model of up and down quarks \cite{manohar};
in a composite Higgs scenario based upon \cite{BHL} this model was considered
by Luty \cite{Luty}):
\bea\label{one}
V_{HY} = g\overline{\Psi}_L\Sigma\Psi_R + h.c.\quad
\text{where}
\quad
\Psi= \left( \begin{array}{c}
t \\ 
b%
\end{array}%
\right).
\eea
In Eq.~(\ref{one}), $\Sigma$ is a $2\times 2$ complex matrix and 
$V_{HY}$ is invariant under $\Sigma \rightarrow U_L \Sigma U^\dagger_R$
with ${\Psi}_L \rightarrow U_L{\Psi}_L$
and ${\Psi}_R \rightarrow U_R{\Psi}_R$.
Note that the $U(1)_Y$ generator $Y$ of the SM now becomes
$Y= I_{3R} + (B-L)/2 $ and furthermore $\Sigma$ is neutral
under $B-L$ which implies its weak hypercharge $[Y,\Sigma]=\Sigma \, I_{3R}$. 

An additional $ U(1)_A$ axial symmetry arises as an overall phase transformation
of $\Sigma\rightarrow e^{i\theta}\Sigma$, accompanied
by ${\Psi}_L\rightarrow e^{i\theta/2}{\Psi}_L$
and ${\Psi}_R\rightarrow e^{-i\theta/2}{\Psi}_R$.  
$\Sigma$ can be written
in terms of two column doublets,
\beq
\label{sigma}
\Sigma=(H_0,H^c_b),
\eeq
where  $H^c=i\sigma_2 H^*$. Under $SU(2)_L$ we have 
$H\rightarrow U_L H$, and 
$H^c\rightarrow U_L H^c$ where we note that the weak hypercharge eigenvalues
of $H$ and $H^c$ have opposite signs.
The HY couplings of Eq.~(\ref{one}) become
\bea
V_{HY} = g_t \left( 
\bar{t}, \, \bar{b}
\right)_{L} H_{0} t_{R} +g_b \left( 
\bar{t}, \, 
\bar{b}%
\right)_{L} H^c_{b} b_{R} + h.c.,
\eea
where the $SU(2)_R$ symmetry has forced $g= g_t= g_b$.

Note that this becomes identical to the SM 
if we make the identification
\beq
\label{four}
H_b \rightarrow \epsilon H_0,\qquad \epsilon = \frac{m_b}{m_t}=0.0234.
\eeq
The HY coupling of the $ b $-quark to $H_0$ is then $y_b =\epsilon g_b$,
but with $\epsilon \ll 1$ the $SU(2)_R$ symmetry is then lost. 
The SM Higgs boson (SMH), $ H_0 $, in the absence of $H_b$, has the usual SM potential:
\bea
\label{potential0}
V_{\text{Higgs}}= M_{0}^{2}H_0^{\dagger }H_0+\frac{\lambda}{2} (H_0^{\dagger }H_0)^2,
\eea
where $M_{0}^{2}\simeq -( 88.4)$ GeV$^2$ (note the sign), and $\lambda \simeq 0.25 $.
Minimizing this, we find that the Higgs field $H_0$  acquires its usual VEV,  
$v=|M_0|/\sqrt{\lambda}=174$ GeV,  
and the observed physical Higgs boson, $h$, acquires
mass $m_h=\sqrt{2} |M_0| \simeq 125$ GeV.
In our present scheme Eqs.~(\ref{four}) and (\ref{potential0}) arise at low energies dynamically.

Given Eqs.~(\ref{one})and (\ref{sigma}), 
we postulate a new potential,
\bea
\label{potential1}
V&=&
M_1^2 Tr\left( \Sigma ^{\dagger}\Sigma \right)
-M_2^2 Tr\left( \Sigma ^{\dagger }\Sigma \sigma_3 \right)
\\ 
&& \!\!\!\!\!\!\!\!\!    \!\!\!\!\!  +\mu ^{2}\left(e^{i\theta} \det\Sigma + h.c.\right)
+ \frac{\lambda_1 }{2} Tr\left( \Sigma ^{\dagger }\Sigma\right)^{2}
+ \lambda_2 |\det\Sigma|^2.
\nonumber 
\eea
We have written the potential in the  $\Sigma$ notation  in order
to  display the symmetries more clearly.
In the above, $\sigma_3$
acts on the $SU(2)_R$ side of $\Sigma$, hence
in the limit $M_2^2 = 0$
the potential $V$ is invariant under $SU(2)_R$, while 
 the $\mu^2$ term breaks the 
additional $U(1)_{A}$. The associated CP-phase
can be removed by field redefinition in the present model.
If the exact $U(1)_A$ symmetry is imposed
on the  $d=4$ terms in the potential then operators such
as,  $e^{i\alpha} (\det\Sigma)Tr\left( \Sigma ^{\dagger }\Sigma \right)$,
$e^{i\alpha'} (\det\Sigma)^2$,  etc., are forbidden.
Noting the identity 
$(Tr( \Sigma^\dagger \Sigma) )^2=Tr( \Sigma^\dagger \Sigma)^2 + 2\det \Sigma^\dagger \Sigma $,
we obtain only the two indicated $d=4 $ terms as the maximal form of the invariant
potential.

The global symmetries of $G$ restrict the $d=4$ terms of this model.
They also make the $d=2$ symmetry breaking terms technically natural.  
For example, the $d=4$ operator $Tr(\Sigma^\dagger \Sigma)\det\Sigma$
is disallowed by $U(1)_A$.  However the  $d=2$  `t Hooft operator, $\mu^2 \det\Sigma$,
breaks $U(1)_A$,
is perturbatively multiplicatively renormalized and is technically naturally small
(we are not considering nonperturbative instantons here).
The analogue
of the Higgs boson mass is the $d=2$ operator, $Tr(\Sigma^\dagger \Sigma)$, which is allowed
by $G$, but is no  less natural than the SM Higgs mass term. 
The other two $d=2$ terms, $\mu^2 \det\Sigma$  
and $Tr(\Sigma^\dagger \Sigma \sigma_z)$,
explicitly break $U(1)_A $ and $SU(2)_R$, respectively,
are multiplicatively renormalized and 
can be technically naturally small.
In fact, our theory is no less natural than the SM
and this is true for the entire scalar democracy!.

Using Eq.~(\ref{sigma}), $V$ can be written in terms of $H_0$ and $H_b$:
\bea
\label{potb}
V &=& 
M_{H}^{2}H_{0}^{\dagger }H_{0}+M_{b}^{2}H_{b}^{\dagger }H_{b}
+\mu^{2}\left( e^{i\theta} H_{0}^{\dagger }H_{b} + h.c. \right) 
\nonumber \\
&& +\frac{\lambda }{2}\left(
H_{0}^{\dagger }H_{0}+H_{b}^{\dagger }H_{b}\right) ^{2}+\lambda ^{\prime
}\left( H_{0}^{\dagger }H_{b}H_{b}^{\dagger }H_{0}\right),
\eea
where $\lambda_2=\lambda' + \lambda$, $M_H^2=M_1^2-M_2^2$, and  $M_b^2=M_1^2+M_2^2$.
In the limit $M_b^2\rightarrow \infty$ the field $H_b$  decouples,
and Eq.~(\ref{potb}) reduces to the SM  if
$M_H^2 \rightarrow M_0^2$, and $\lambda \rightarrow 0.25$ thereby recovering Eq.~(\ref{potential0}).

We assume the quartic couplings are all of order the SM  value $\lambda\sim 0.25$.
{ Those associated with the new heavy Higgs bosons, such as $\lambda'\sim \lambda$, 
will therefore contribute negligibly small effects since they involve 
large, positive, $M^2$ heavy Higgs fields. We also set $\theta=0$.} 

Then, varying the potential
with respect to $H_b$, 
the low momentum components of $H_b$ become locked to $H_0$:
\beq
H_{b}= -\frac{\mu^2}{M_b^2}  H_{0}+ O(\lambda, \lambda').
\eeq
Substituting back into $V$ we recover the SMH potential
\bea
V &=& 
 M_0^2 H_{0}^{\dagger }H_{0}
 +\frac{\lambda }{2}\left(
H_{0}^{\dagger }H_{0}\right)^{2}+O\left(\frac{\mu^2}{M_b^2}\right),
\eea
with
\beq
\label{M0}
M_0^2= M_{H}^{2}-\frac{\mu^4}{M_b^2}.
\eeq
Note that, even with $M_H^2$ positive, $M_0^2=-(88.4)^2$ GeV$^2$
can be driven to its negative value by
the mixing with $H_b$ (level repulsion).
We minimize the SMH potential and define 
\bea
\label{SMHpot}
H_{0}=\left( 
\begin{array}{c}
v +\frac{1}{\sqrt{2}}h\\ 
0
\end{array}%
\right), \qquad v=  {174}\;\makebox{GeV},
\eea
in the unitary gauge. 
The minimum of Eq.~(\ref{SMHpot}) 
yields the usual SM result
\beq
v^2 = -M_0^2/\lambda, \qquad m_h  = \sqrt{2}|M_0|=125\;\makebox{GeV},
\eeq
where $m_h$ is the propagating Higgs boson mass.
We can then write
\beq
H_b\rightarrow H_b -\frac{\mu^2}{M_b^2}\left( 
\begin{array}{c}
v +\frac{1}{\sqrt{2}}h\\ 
0
\end{array}%
\right),
\eeq
for the full $H_b$ field. 
This is a  linearized (small angle) approximation to the mixing,
and si reasonably insensitive to the small $\lambda, \lambda'\ll1$.

Note  the effect of ``level repulsion'' of the Higgs mass, $M_0^2$,
downward due to the mixing with heavier $H_b$.
The level repulsion in  the presence of $\mu^2$ and $M_b^2$
occurs due to an approximate ``seesaw'' Higgs mass matrix
\beq
\label{matrix}
\begin{pmatrix}
	M_H^2 & \mu^2 \\
	\mu^2 & M_b^2
\end{pmatrix}.
\eeq
The input value of the mass term $M_H^2$ is
unknown and in principle arbitrary, and can 
have either sign. 
We can presumably bound $M_H^2 \gtrsim  {1}{GeV^2}$ from below, 
since QCD effects will mix glueballs and the QCD-$\sigma$-meson with $H_0$ in this limit.
As $M_H^2$ is otherwise arbitrary,
we might then expect that the most probable
value is $M_H^2\sim  {1}{GeV^2}$, and $|M_H^2|\ll(\mu^2, M_b^2)$. 

Let
us consider the case  $M_b^2>>M_H^2$.
Then  Eq.~(\ref{matrix}) has eigenvalues
$M_0^2=-\mu^4/M_b^2$, and $ M_b^2$.
Thus, in the limit of small, nonzero $|M_H^2|$,
we see that a negative $M_0^2$ arises naturally, and to a good approximation the physical Higgs mass 
is generated entirely by this negative mixing term.

The mass mixing causes the neutral component of $H_b$ to acquire a small VEV (``tadpole'') 
of  $-v(\mu^2/M_b^2)$.  This implies that the SM HY-coupling of the $b$-quark is
induced with the small value $y_b =g_b(\mu^2/M_b^2)$
(note $g_b$ is positive with a phase redefinition of $b_R$).
The $b$-quark then receives its mass from $H_{b}$,
\beq
\label{mub1}
m_b =  g_b(m_b)v\frac{\mu^{2}}{M_b^{2}}
=m_{t}\frac{g_b(m_b)\mu^{2}}{g_t(m_t)M_b^{2}},
\eeq
where we have indicated the renormalization group (RG) scales 
at which these couplings should be evaluated.

In a larger scalar democracy framework 
both $g_t(m)$ and $g_b(m)$ have 
a common renormalization group equation modulo $U(1)_Y$ effects
This implies 
$g(M_P)= g_t(M_P)=g_b(M_P) \gta 1$. Then,
we will predict 
$g(M_b)\simeq  g_t(M_b)\simeq g_b(M_b)$,  eg., where the values 
at the mass scale $M_b$ are determined by the RG fixed point.  

Furthermore, we find that $g_t(m)$, and moreso $g_b(m)$, increase somewhat as we evolve 
downward from $M_b$ to $m_t$ or $m_b$. The top quark mass is then $m_t=g_t(m_t) v$ 
where $v$ is the SM Higgs VEV.  From these effects we obtain
the ratio
\beq
\label{Rtau}
R_b = \frac{g_b(m_b)}{g_t(m_t)} \simeq 1.5.
\eeq
The $b$-quark then receives its mass from the tadpole VEV of $H_{b}$.
\beq 
\label{mub}
m_b = g_b(m_b) v\frac{\mu^{2}}{M_b^{2}}=m_{t} R_b\frac{\mu^{2}}{M_b^{2}}.
\eeq
In the case that the Higgs mass, $M_0^2$,
is due entirely to the level repulsion by $H_b$,  ie. $M_H^2 =0$, 
(see also \cite{Ishida} and references therein)
using Eqs.~(\ref{M0}), (\ref{Rtau}) and (\ref{mub}), we obtain
a predicted mass of the $H_b$, 
\beq
M_b = \frac{m_{t}}{m_b} R_b|M_0|  \simeq   {5.5}\;\makebox{TeV},
\eeq
with $m_b =  {4.18}\;\makebox{GeV} $, $ m_t= {173}\;\makebox{GeV}$, and $|M_0| =  {88.4}\;\makebox{GeV}$.
We remind the reader we have
 ignored the effects of the quartic couplings $\lambda$,
which we expect are small.
Moreover, the quartic couplings do not enter the mixing, because
terms such as, $H^\dagger_0 H_0 H^\dagger_0 H_b $ are forbidden
by our symmetry.  The remaining terms only act as slight
shifts in the masses, never larger than $\sim \lambda v^2$
and can be safely ignored.

This is a key prediction of the model.  In fact, we can argue
that with $M_H^2$ nonzero, but with small fine tuning (see below), the result
$M_b\lta  {5.5}\;\makebox{TeV}$ is obtained.
This mass scale is accessible to the LHC
with luminosity and energy upgrades, and we feel represents an important
target for discovery of the first sequential Higgs Boson.

The simple $(t,b)$ system described above can be extended to the third generation
leptons $(\nu_\tau, \tau)$. 
Remarkably the predictions for the mass spectrum
are sensitive to the mechanism of neutrino mass generation.
The next sequential massive Higgs iso-doublet, in addition
to $H_b$, is likely to include the $H_\tau$,
and possibly also $H_\nu$, which is dependent
upon whether neutrino masses are Majorana
or Dirac in nature.   If neutrino masses are Dirac
then $H_\nu$ is very heavy, $\sim 10^{16}$ GeV, and then
have a Dirac seesaw and we can ignore $H_\nu$.
The details of this are given in \cite{HMTT,HMTT2}.

Hence, the $\tau$ mass is given analogously by
\beq
\label{mutau}
m_\tau =m_{t} R_\tau \frac{\mu_2^{2}}{M_\tau^{2}}.
\eeq
and we expect $R_\tau\sim 0.7$. 
$H_\tau$ and $H_b$ now simultaneously contribute to the SMH mass
\beq
M_{0}^2 =  M_H^2-\frac{\mu^{4}}{M_b^{2}}-\frac{\mu_2^{4}}{M_\tau^{2}}.
\eeq
(we remind the reader that $M_0^2=-(88.4)^2$ GeV$^2$ is negative
as defined in Eq.(\ref{potential0})).
Using Eqs.~(\ref{mub1}) and (\ref{mutau}) this yields an elliptical 
constraint on the heavy Higgs masses $M_b$ and $M_\tau$,
\beq
\label{ellipse}
|M_{0}|^2 + M_H^2=\frac{m_b^2M_b^{2}}{m_t^2R_b^2}+\frac{m_\tau^2 M_\tau^{2}}{m^2_t R_\tau^2},
\eeq
for fixed $ M_H^2 $. Bear in mind that the Higgs potential input mass, $M_H^2$,
is {\em a priori} unkown,
while $M_0^2= -( {88.4}\;\makebox{GeV})^2 $ is known from the Higgs boson mass, 
$m_h=\sqrt{2}|M_0|=125$ GeV.

If we make the assumption $M_H^2=0$ 
the ellipse is shown as  the red-dashed line in Fig.(5).
If we further assume both $H_b$ 
and $H_\tau$ contribute equally
to the SMH mass, then we obtain from Eq.~(\ref{ellipse})
\beq
 M_b\simeq  {3.6}\;\makebox{TeV}, \qquad   M_\tau \simeq  {4.2}\;\makebox{TeV}.
\eeq
Of course, we can raise (lower) these masses by introducing the
bare positive (negative) $M_H^2$.  However, we do not
want to excessively fine tune the difference $-|M_{0}^2| + M_H^2$.

As an alternative way of estimating the Higgs masses we follow 
the same procedure for estimating the fine tuning
as is sometimes used in the MSSM or composite Higgs models.

\begin{figure}[t!]
	\includegraphics[width=0.5\textwidth]{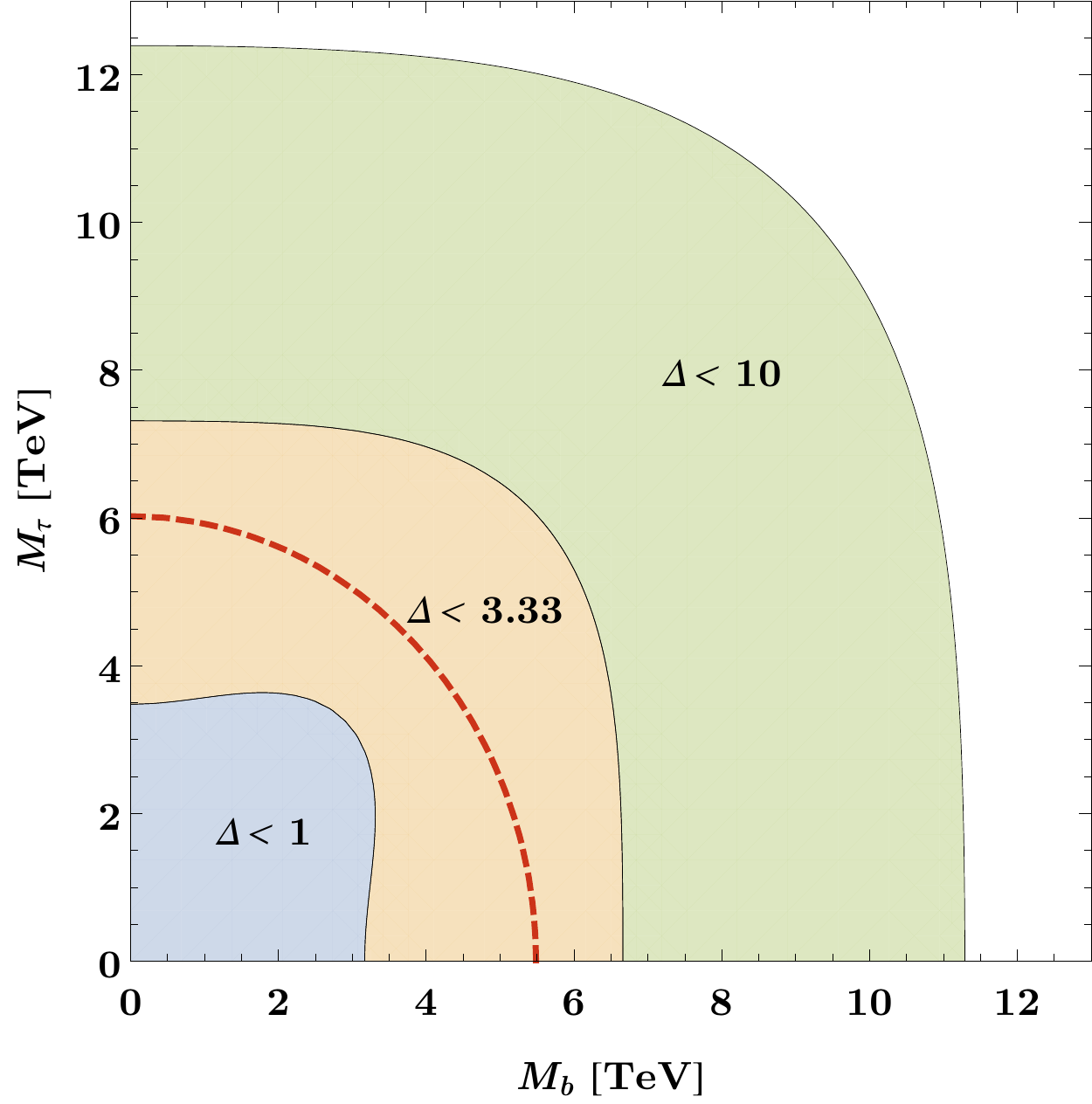}
	\caption{The fine tuning associated with different 
	values of the $ M_b $ and $ M_\tau $ parameters. The 
	remaining three parameters $ \mu $, $ \mu_\tau $, 
	and $ M_H $ are fixed by the physical choices of 
	fermion masses and Higgs VEV. The red dashed line 
	corresponds to $ M^2_H = 0 $, and the origin to $M_H^2 =-|M_0^2|$.}
\end{figure}
\begin{figure*}[t!]
	\includegraphics[width=1.0\textwidth]{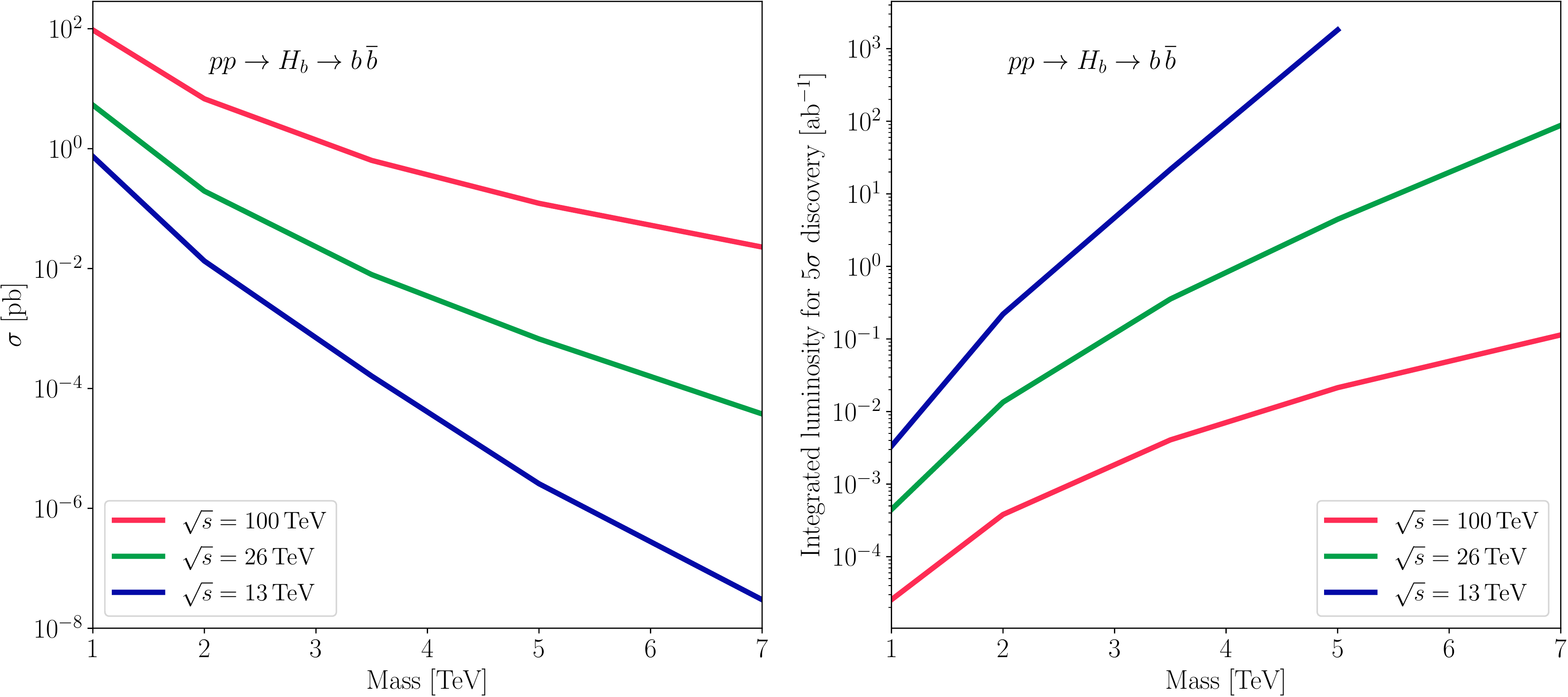}
	\caption{The left plot shows the estimated cross-sections for $pp\rightarrow X + H_b(\rightarrow \bar{b}b)$
	at the LHC $13$ TeV,  LHC upgraded $26$ TeV, and a $100$ TeV $pp$ collider. The 
	cross sections where calculated using structure functions with perturbative $b$ and $\bar{b}$, 
	applying a $100$ GeV $p_T$ cut  on the
	final state $b$-jets, using \texttt{MadGraph5\_aMC@NLO} and \texttt{CalcHEP}.
	The right plot displays the estimated $5\sigma$ integrated luminosites for $H_b$ in the $\bar{b}b$
	final state, applying a Breit-Wigner as described in the text, 
	and  assuming a $50\%$ double $b$-tagging efficiency.}
	\label{plot1}
\end{figure*}

Examining just the theory with $ H_b $ alone, we can  
combine the log-derivative sensitivities of $M_0^2$, $m_b$
to three input parameters $M_H^2$, $\mu^2_b$, and
$M_b^2$. Then we find the 
``no fine tuning limit'' 
($ \Delta \leq 1 $) translates into the 
restrictive constraint 
	\begin{equation}
	 M_b \leq   {3.1}\;\makebox{TeV}.
	\end{equation}	
 The lightest values
of $M_b$ (and $M_\tau$) correspond to $-|M_0^2| = M_H^2$
with no contribution from $M_b^2$  ($M_\tau^2$). 
Indeed, with more heavy Higgses we have generically smaller masses and 
these states become
more easily accessible to the existing LHC.

We also consider the conventional possibility that
the neutrino has a Majorana mass term, which is an extension of the
physics beyond the minimal model.arable to $m_\tau$.
We can then suppress the physical
neutrino mass, $m_\nu$, by allowing a the usual Type I seesaw.
We  thus postulate a large Majorana mass term for the
ungauged $\nu_R$:
\beq
M\bar{\nu}^c_R\nu_R + h.c.
\eeq
Integrating out $\nu_R$
we then have an induced $d=5$ operator that
generates a Majorana mass term 
for the left-handed neutrinos 
through the VEV of $H_\nu$,
\bea
V_M= \frac{g^2}{2M}\bar{\Psi}_L\Sigma_\ell\Sigma^{c\dagger}_\ell \Psi^c_L + h.c.
\eea

In this scenario there is no restriction that requires the $H_\nu$ mass
to be heavy, and the neutrino physical mass is now small, given by $m_\nu \sim m_{D\nu}^2/M
\sim m_\tau^2/M$
where $m_{D\nu}\sim m_\tau$ is the neutrino Dirac mass. $H_\nu$
acquires a  VEV by mixing, and has comparable 
feedback on the Higgs mass as $H_\tau$,
\beq
H_\nu= -\frac{\mu_1^2}{M_{\nu}^{2}}H_0,\qquad \;\; \delta M_0^2
= -\frac{\mu_1^4}{M_{\nu}^{2}}.
\eeq
We now have an ellipsoid for the masses
$M_b$, $M_\tau$ and $M_\nu$
\beq
|M_{0}|^2 + M_H^2=\frac{m_b^2M_b^{2}}{m_t^2R_b^2}+\frac{m_\tau^2 M_\tau^{2}}{m^2_t R_\tau^2}
+\frac{m_{D \nu}^2 M_\nu^{2}}{m^2_t R_\nu^2}.
\eeq 
With additional light Higgs fields, the elliptical constraint forces all 
of the Higgs masses to smaller values.   We emphasize that
the mass bounds of Fig.~(\ref{fig:Ft_regions}) should
be viewed as upper limits on the Higgs mass spectrum that could
be explored at the LHC.

\section{Phenomenology of the Sequential Higgs Bosons}

Presently we  touch upon the collider phenomenology of this model
and refer the reader to \cite{HMTT} for further discussion.
{Our estimated discovery luminosites
for $h^0_b $ are seen to be attainable at the LHC or its upgrades, and we thus encourage
and plan
more detailed studies. Moreover, the lower mass range $\lesssim 1$ TeV
is currently within range of the LHC and collaborations should attemtp to place
limits.}

$H_b$ is an iso-doublet with neutral $h^0_b$  and charged $h^\pm_b$
{\em complex field} components
coupling to $(t,b)$ as  $g_b h^0_b \bar{b}_L b_R + \mathrm{h.c.}$ and $g_b h_b^+ \bar{t}_L b_R + h.c.$
with $g_b \simeq 1$.
At the LHC the $h_b^0$ is singly produced in $pp$ by the perturbative (``intrinsic'')  $b$ and $\bar{b}$
components of the proton,
\beq
pp (b\bar{b}) \rightarrow h^0_{b} \rightarrow b\bar{b}.
\eeq
{Structure functions that contain the $b+\bar{b}$ components are 
available in \texttt{MadGraph5\_aMC@NLO} and \texttt{CalcHEP},
and results obtained by these two simulators are found to be consistent. The resulting
cross-sections for $13$ TeV, $26$ TeV and $100$ TeV are given for various
$M_b$ in \cite{HMTT} and in Fig.~(\ref{plot1}). }

The cross-section
for production at the LHC of $h_b^0$ is $\sigma(h^0_b\rightarrow b\bar{b})
\sim  10^{-4}{pb}$ at $13$ TeV  (or $\sigma(h^0_b\rightarrow b\bar{b}) \sim  10^{-2}{pb}$
at $26$ TeV) for a mass of $M_b ={3.5}$ TeV.
The decay width of $h_b^0$   is large, $\Gamma = 3M_b/16\pi \simeq  {210}$ GeV, for $M_{b}=3.5$ TeV.
To reduce the backgrounds we  
impose a $ {100}\;\makebox{GeV}$ $p_T $ cut on the each $b$ jet
in the quoted cross-sections. Note that the charged  $h_b^+ $ would be produced in association with
$\bar{t}b$, has a significantly smaller cross-section
and we have not analyzed it. 

{We note that
the $h^0$ is centrally produced in a narrow range of rapidity, $|\eta|<1$
which may afford useful cuts, though this somewhat redundant to the $100$ GeV $p_T$
we used in this study. The main backgrounds are high mass $b$-quark-dijet production,
$pp\rightarrow \bar{b}b$ and 
ordinary flavor dijets $pp\rightarrow \bar{q}q$ that fake $\bar{b}b$.
The $pp\rightarrow \bar{b}b$ is mostly forward and requires the $p_T$ cut, 
chosen to be $100$ GeV$/c$. We have not extensively studied
the optimization of this choice for this cut. 
We use the same simulators to generate the backgrounds. 
See \cite{HMTT,HMTT2} for more details.

We estimate that a $5\sigma$ excess in $S/\sqrt{B}$,
in bins spanning twice the full-width of the Breit-Wigner, requires an
integrated luminosity or $M_b=3.5$ TeV of $\sim {20}{ab^{-1}} $
at $13$ TeV, or
$\sim {100}{fb^{-1}} $ at $26$ TeV.\footnote{ ${1}\;{ab} = {1}$ {attobarn} = $10^{-3}$ fb (femtobarn).}  
Similarly, at a mass of $ 5$ TeV one requires 
$\sim  {3}{ab^{-1}}$ at $26$ TeV for a $5\sigma$ discovery.
This assumes double $b$-tagging efficiencies of order $50\%$.
Here the background is assumed to be mainly $gg\rightarrow \bar{b}b$, but the large fake
rate from $gg\rightarrow \bar{q}q$ must be reduced to $\lta 3\%$ for this to apply.
 
Hence according to our estimates, while meaningful bounds
are acheiveable at the current $13$ TeV LHC, particularly $M_b\lta 3.5$ TeV,
the energy doubler is certainly favored for this physics and could cover
the full mass range up to  $\sim 7$ TeV. 

Remarkably, the  $h^0_\tau$ (and $h^0_\nu$), neutral 
components of the associated
iso-doublets,   $H_\tau$ and $H_\nu$, may also be singly
produced  because they can mix with 
$H_b$.
This implies a total cross section
\beq
\sigma(h^0_\tau,h^0_\nu) \sim \theta^2 \sigma(h^0_b),
\eeq
for the mixing angle, $\theta$, between either the states $h^0_\tau$ or $h^0_\nu$ and $h^0_b$. 
$\theta$ could be large, $ \theta \sim 0.3$.
The $h^0_\tau\rightarrow \bar{\tau}\tau$ is visible with $\tau$-tagging, and
the background is also slightly suppressed since the peak is narrower by a factor
of $(g_\tau^2/3g_b^2)\sim0.16$.  Hence at the $26$ TeV LHC discovery is
in principle possible for a $3.5$ TeV state with integrated luminosity
of order $2$ ab$^{-1}$.   

We note that the LHC now has the capability of ruling out
an $H_b$ with $g_b\sim 1$ of mass $\sim 1$ TeV, with current
integrated luminosities, $\sim  {200}{fb^{-1}} $. 


\section{Summary}

The idea that there is a lone Higgs boson in the world  is, to me, absurd.\footnote{We are motivated by the ``one dead mouse'' conjecture
whereby, upon finding one mouse corpse in the closet, there will be many mice living in the walls.}
We count 1176 possible ``slots'' in the SM into which we can insert an s-wave complex scalar
boundstate. Surely nature must make use of these in a more complete way? 

It is certainly possible that the  fermion generations
arise from, e.g., a topological defect, such as a flux-tube in $D=6$ with
$3$ units of flux, or  conifold singularies on Calabi-Yau manifolds.
In this case, perhaps the $SU(48)$ is reduced to $SU(16)$
and a more manageable spectrum of Higgses emerges. 
However, if the fermion generations emerge before a  compositeness scale
of the scalars, then our present scenario may be relevant.

The extended Higgs scalar theory, which we dub ``Scalar Democracy'' is a
composite scalar theory and is, in a sense, an grand expansion of
the old top condensation ideas. Indeed, this realizes
the SM Higgs boson as a $\bar{t}t$ system, and brings in a large number
of testable predictions as we have decribed. 

We haven't discussed the binding mechanism in any detail and we presently have
little to say.  It may be that a $TT$ (stress-tensor)$^2$ effectve NJL model can
emerge (however, this does not seem to offer a large-$N$ binding limit.)
However, I think a nonperturbative gravitational binding may exist
and have proposed (with Gabriella Barenboim) a  model of ``holographic
binding'' in a nontrivial black hole vacuum \cite{Baren}.  In an $SU(6)\times SU(6)$
model of the quark sector we can engineer a large-N model in analogy to topcolor
\cite{HillThomsen}.

In brief, the theory achieves the following:

\begin{enumerate}

\item Every $s$-wave fermion bilinear is bound at the Planck scale to form a complex
scalar field with a universal coupling $g$.  This implies 1176 
composite complex scalar fields, and a maximal sequential Higgs doublet spectrum.

\item The masses and mixings of the Higgs fields
are introduced
as input parameters which  can explain the spectrum of SM fermions
and their CKM masses and mixings while maintaining consistency with rare weak processes.  

\item The extended mass spectrum
is no less technically natural than
the SM since most of these terms have custodial symmetries.
These $d=2$ explicit mass terms might be replaced by new $d=4$ quartic interactions in an extension of the theory.

\item   The theory predicts $m_{top}$ through the quasi-fixed point.

\item   The standard model Higgs boson is a $\bar{t}t$ composite.
        
 \item  There is one Universal Higgs-Yukawa coupling: $g = g_{top}$ for quarks (with RG effects
        $g \approx 0.7 g_{top}$ for leptons).
 
 \item  CKM physics is no longer HY texture; rather it is determined by the spectrum of masses and mixings of
        maximally extended Higgs sector.  
 
 \item  In the simplest $t$-$b$ subsector we explain the $b$-quark mass, or $y_b=0.024$, by the existence
         the new isodoublet $H_b$ with mass $\lta 5.5$ TeV. This is
         detectable $< 3.5$ TeV at LHC;  $<5.5$ TeV at the energy doubled LHC; $\sim 30$ TeV at $100$ TeV machine.
         
 \item  The LHC can already place useful limits on the $H_b$ and $H_\tau$ masses.
 
 \item   The theory explains the tachyonic (negative mass$^2$) SM Higgs boson by a seesaw-like
         mixing.

\end{enumerate}

Our main goal is to have a successful theory of $m_{top}$, $m_{Higgs}$ 
based upon the renormalization group, and any useful predictions
of sequential Higgs bosons such as $H_b$, $H_\tau$ etc. that may
be addressable in current (LHC) or future machines. We will elaborate these
issues further elsewhere \cite{HillThomsen}.

\section*{Acknowledgments}

I acknowledge
the  Fermi Research Alliance, LLC under Contract No.~DE-AC02-07CH11359 with the U.S.~Department of Energy,  Office of Science, Office of High Energy Physics.
I thank the various particpants of the Simplicity III workshop for 
useful discussions, in particular Lena Funcke,
Gerard 't Hooft, and Neil Turok.  The present summary describes work in
progress with Anders E. Thomsen.

\end{document}